\documentclass[11pt,a4paper]{article}

\pdfoutput=1

\usepackage{graphicx}
\usepackage{comment}
\usepackage{epstopdf}
\usepackage{jcappub}
\usepackage{amsmath,amsthm,latexsym,amssymb,amsfonts}
\usepackage{fontenc,dsfont,layout}
\usepackage{hyperref}
\usepackage{psfrag}
\usepackage[cp1250]{inputenc}
\usepackage[usenames,dvipsnames]{xcolor}

\numberwithin{equation}{section}

\title{Scalar-tensor linear inflation}

\author[a]{Micha{\l}~Artymowski}
\author[b]{Antonio Racioppi}

\affiliation[a]{Institute of Physics, Jagiellonian University\\
{\L}ojasiewicza 11, 30-348 Krak{\'o}w, Poland}
\affiliation[b]{National Institute of Chemical Physics and Biophysics, R\"avala 10, 10143 Tallinn, Estonia}

\emailAdd{Michal.Artymowski@uj.edu.pl}
\emailAdd{Antonio.Racioppi@kbfi.ee}

\abstract{We investigate two approaches to non-minimally coupled gravity theories which present linear inflation as attractor solution: a) the scalar-tensor theory approach, where we look for a scalar-tensor theory that would restore results of linear inflation in the strong coupling limit for a non-minimal coupling to gravity of the form of $f(\varphi)R/2$; b) the particle physics approach, where we motivate the form of the Jordan frame potential by loop corrections to the inflaton field. In both cases the Jordan frame potentials are modifications of the induced gravity inflationary scenario, but instead of the Starobinsky attractor they lead to linear inflation in the strong coupling limit.}

\keywords{Inflation, scalar-tensor theory, loop corrections}

\begin{document}
\maketitle

\section{Introduction} \label{sec:Introduction}

Cosmic inflation \cite{Lyth:1998xn,Liddle:2000dt,Mazumdar:2010sa} is a well established theory of the early universe with good consistency with data \cite{Ade:2015lrj}. It predicts the accelerated expansion of space together with flat power spectrum of primordial inhomogeneities of cosmic microwave background, which is measured by several experiments. The latest data from PLANCK \cite{Ade:2015lrj} and BICEP \cite{Array:2015xqh} experiments put stronger constraints on the tensor-to-scalar ratio $r$, which tells us about the amplitude of primordial gravitational waves and about the scale of inflation. The combined results of those experiments \cite{Array:2015xqh} shows that the linear inflation predictions for $r$ in function of the scalar spectral index ($n_s$)
lie on the very edge of the $2 \sigma$ limit of the best fit to the data.\footnote{Linear inflation predicts $(n_s,r) = (1-3N_\star/2,4/N_\star)$, where $N_\star$ is the number of e-foldings at the freeze out of the pivot scale.} This means that besides of the theoretical motivation of linear inflation, this class of models can be confirmed or ruled out quite soon.
\\*

In this paper we are combining two approaches to analyse attractor solutions that lead to linear inflation. On the one hand side we use the effective field theory approach in the Scalar-Tensor (ST) theory frame. Namely we ask for the conditions of the Jordan frame (JF) potential, which would lead to the Einstein frame (EF) linear potential in certain limit of the parameter space, i.e. in the strong coupling limit. This approach is also motivated by the induced gravity inflationary scenario, in which the Starobinsky-like plateau is obtained for the wide class of scalar-tensor theories \cite{Kallosh:2013tua,Kallosh:2014laa,Giudice:2014toa,Kallosh:2014rha}. On the other hand side we motivate our work by quantum field theory. It is proven that loop corrections to inflationary potentials may play a relevant role \cite{Kannike:2014mia,Marzola:2015xbh,Marzola:2016xgb}, allowing the dynamical generation of the Planck scale \cite{Kannike:2015apa,Kannike:2015kda}, predicting super-heavy dark matter \cite{Farzinnia:2015fka,Kannike:2016jfs} and, in presence of a non-minimal coupling to gravity, leading to linear inflation predictions \cite{Kannike:2015kda,Rinaldi:2015yoa,Barrie:2016rnv}. This gives a particle physics point of view to our work and provides an additional motivation to the results obtained in the effective field theory approach. In this work we also investigate under which circumstances those two approaches can meet.
\\*

The structure of this paper is as follows. In Sec. \ref{sec:general} we obtain a particular form of the scalar-tensor theory, which in the strong coupling limit gives the linear Einstein frame potential. In Sec. \ref{sec:QFT} we present the particle physics motivation for the model. Finally we conclude in Sec. \ref{sec:Summary}.


\section{Linear inflation for general form of non-minimal coupling} \label{sec:general}

\subsection{The Scalar-Tensor theory approach}

Let us consider the following action of the scalar-tensor theory with the flat FRW metric tensor
\begin{equation}
S = \int d^4x \sqrt{-g}\left(\frac{M_P^2}{2}f(\varphi)R - \frac{1}{2}K(\varphi)(\partial \varphi)^2\ - U(\varphi)\right) \, .
\label{eq:JframeL}
\end{equation}
By a field redefinition one can always set $K(\varphi) = 1$, which we will assume in the further parts of this work. {Note that $K(\varphi) \neq 1$ would significantly change formulas written below and for given $f(\varphi)$ it would also change predictions of the theory.} In the $f \to 1$ limit one restores general relativity (GR). In order to avoid repulsive gravity one requires $f(\varphi) > 0$. The Einstein frame metric tensor, field and scalar potential are equal to
\begin{equation}
\tilde{g}_{\mu\nu} = f(\varphi) g_{\mu\nu} \, , \qquad \frac{d\phi}{d\varphi} = \sqrt{\frac{3}{2}\left(M_P\frac{f_\varphi}{f}\right)^2+\frac{1}{f}} \, , \qquad  V =  \frac{U}{f^2} \label{eq:EF}
\end{equation}
respectively, {where $f_\varphi := \frac{df}{d\varphi}$}. Furthermore, the Einstein frame action is the following
\begin{equation}
S = \int d^4 x \sqrt{-\tilde{g}}\left(\frac{M_P^2}{2}\tilde{R} - \frac{1}{2}(\partial \phi)^2\ - V(\phi)\right) \, .
\end{equation}
Note that in the EF the scalaron is minimally coupled to the EF metric tensor, just like in the GR frame. In the $M_P^2 f_\varphi^2 \gg f$ limit (which we will denote as the strong coupling limit) one obtains
\begin{equation}
\phi \simeq \sqrt{\frac{3}{2}} M_P \log f + C \, , \label{eq:StrongCoupling}
\end{equation}
{ where $C$ is a constant of integration that does not contain any physical information,  since physical quantities, like slow-roll parameters (see eq. (\ref{eq:slowroll})), are functions of field-value derivatives and therefore insensitive to the actual value of $C$. According to the convenience of the case, common choices are $C = - \sqrt{\frac{3}{2}}M_P\log f(\varphi_0)$, where $\varphi_0$ defines the value when $\varphi$ enters the strong coupling regime, or $C=0$.
}
\\*

{The Einstein frame approach enables us to calculate the inflationary dynamics and power spectra of primordial inhomogeneities within the minimally coupled model. Following this approach one finds
\begin{equation}
\tilde{N} \simeq \frac{1}{M_P^2} \int\frac{V}{V_\phi} d \phi \, ,
\end{equation}
where $\tilde{N}$ is the number of e-folds in the Einstein frame. Having $\tilde{N}(\phi)$ one can define $\phi_\star = \phi(N_\star)$, where $N_\star$ is a number of e-folds which corresponds to the horizon exit of the pivot scale. Since we do not know exactly thermal history of the Universe the value of $N_\star$ is uncertain. We choose $N_\star = 60$, which is one of the most popular choices in inflationary physics around the GUT scale. All of the power spectra of primordial inhomogeneities obtained in this work are calculated within the slow-roll approximation in the Einstein frame. After the normalisation of inhomogeneities at the pivot scale we calculate tensor-to-scalar ratio $r$ and a spectral index $n_s$ defined by
\begin{equation}
r =  \left. 8M_P^2\left(\frac{V_\phi}{V}\right)^2 \right |_{\phi = \phi_\star} \, , \qquad n_s = 1 - M_P^2\left.\left(3 \left(\frac{V_\phi}{V}\right)^2 - 2\frac{V_{\phi\phi}}{V} \right)\right |_{\phi = \phi_\star}\,  ,
\label{eq:slowroll}
\end{equation}
which we will compare with constraints from PLANCK/BICEP experiments.}
\\*

\begin{figure}
\centering
\includegraphics[height=5.7cm,bb=0 0 360 297
]{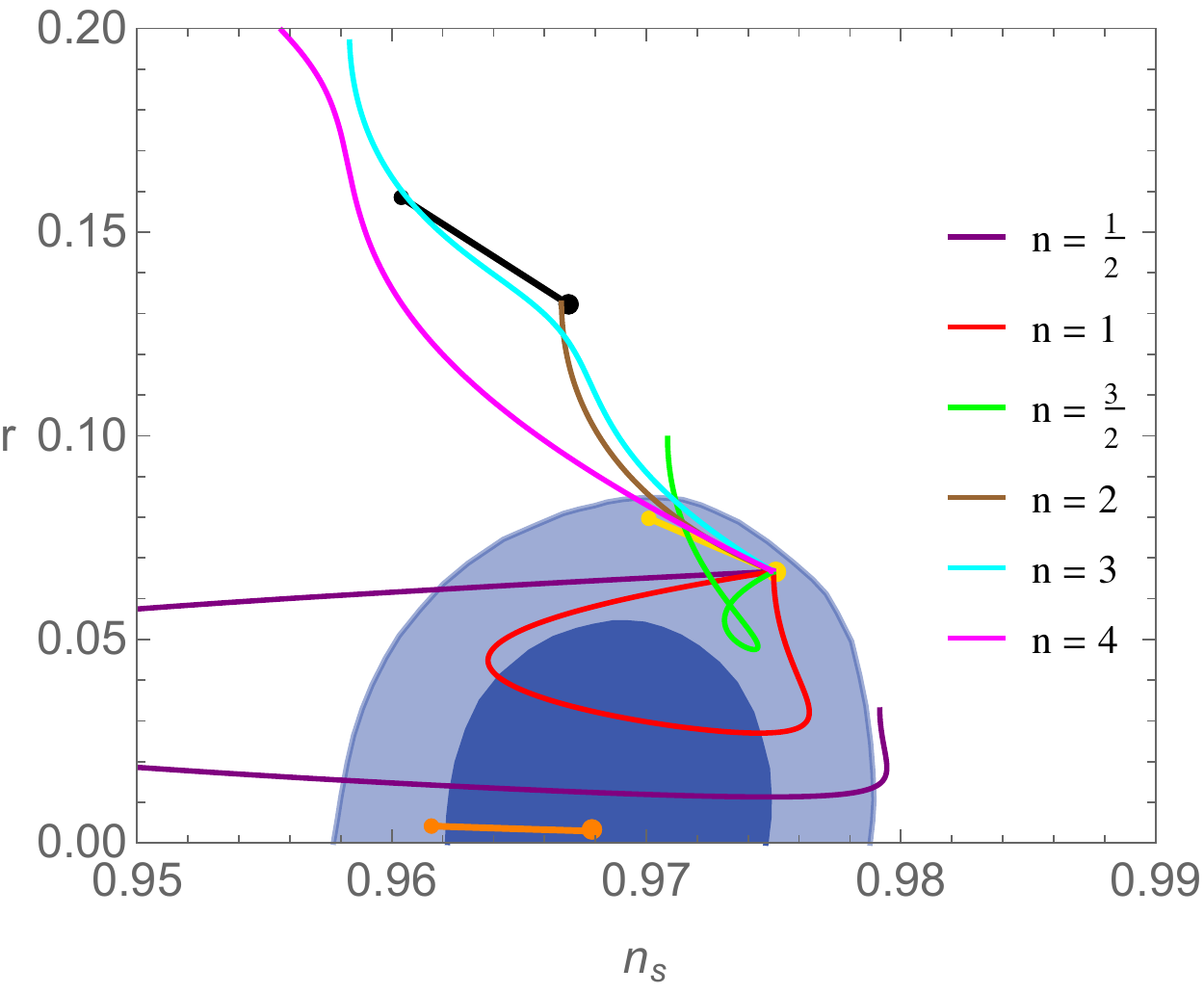}
\hspace{0.5cm}
\includegraphics[height=5.7cm,bb=0 0 360 295
]{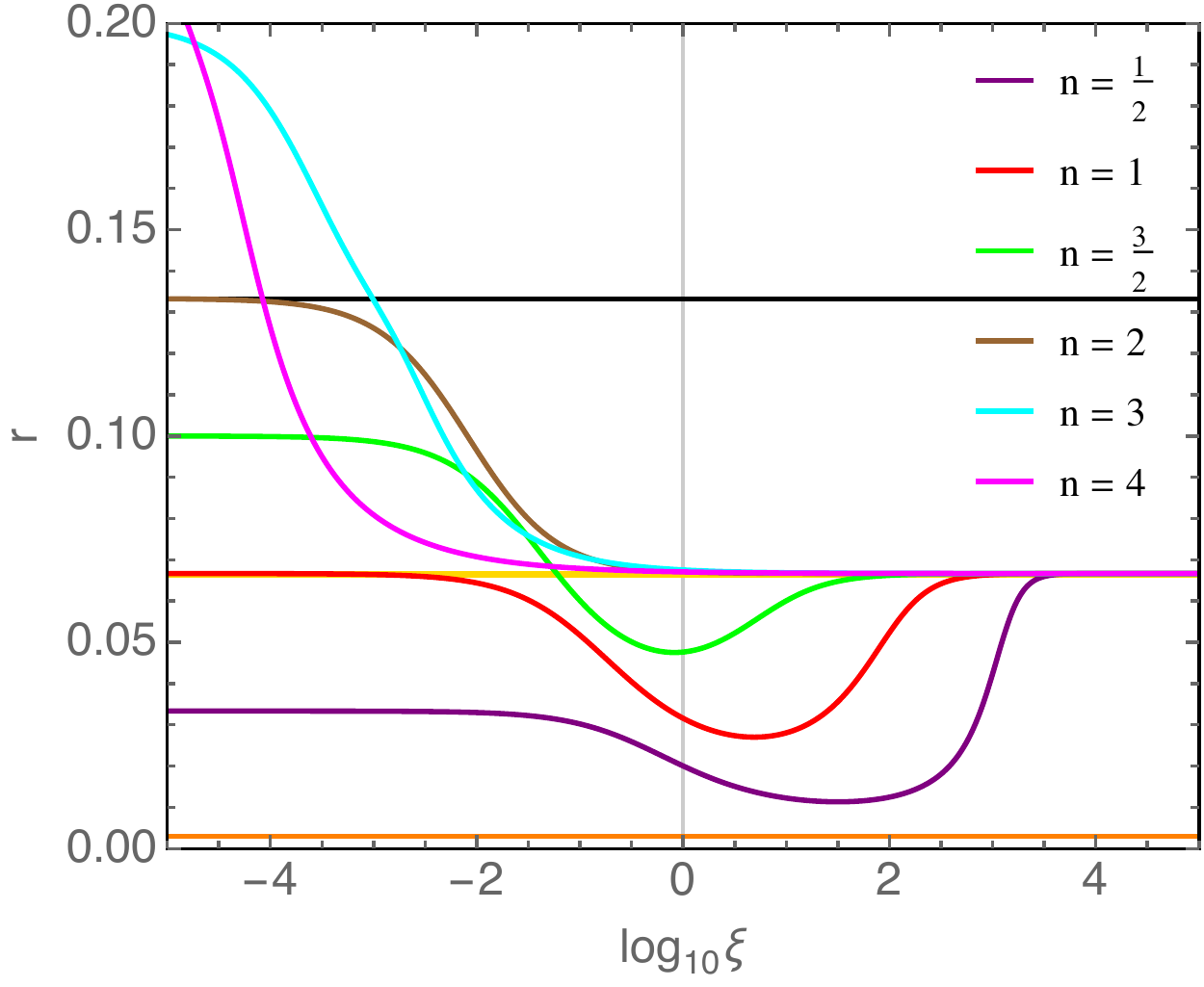}
\caption{Results of for the (\ref{eq:U1}) model with $f(\varphi) = 1 + \xi (\varphi/M_P)^{n}$ . All solutions meet in the attractor point of linear inflation. {The $n=1$ case fits to the linear inflation data also for small values of $\xi$, since the (\ref{eq:U2}) model for $n=1$ gives linear inflation in weak coupling regime too.}
}
\label{fig:rvsnST2}

\end{figure}

\begin{figure}
\centering
\includegraphics[height=5.7cm,bb=0 0 360 297
]{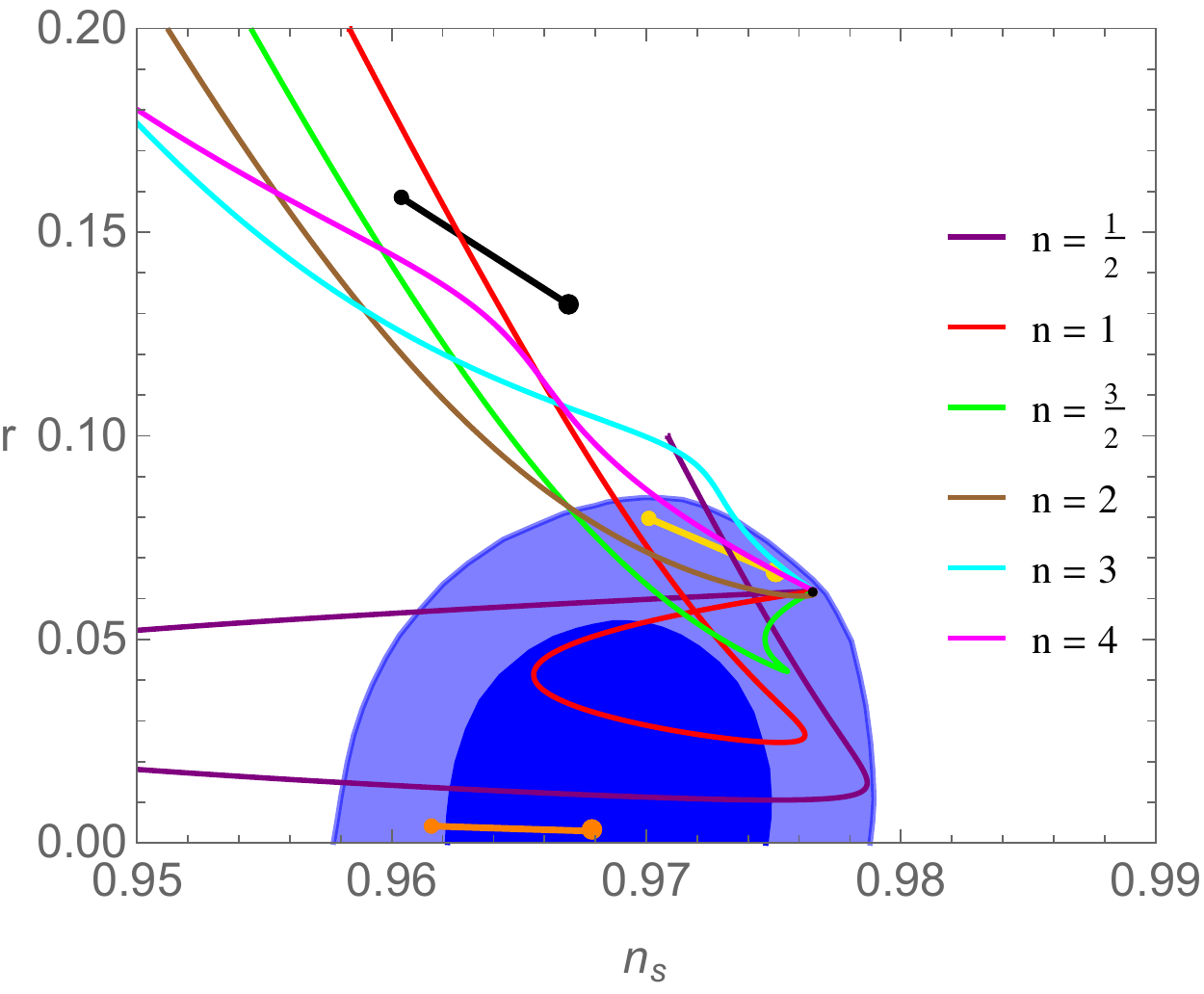}
\hspace{0.5cm}
\includegraphics[height=5.7cm,bb=0 0 360 295
]{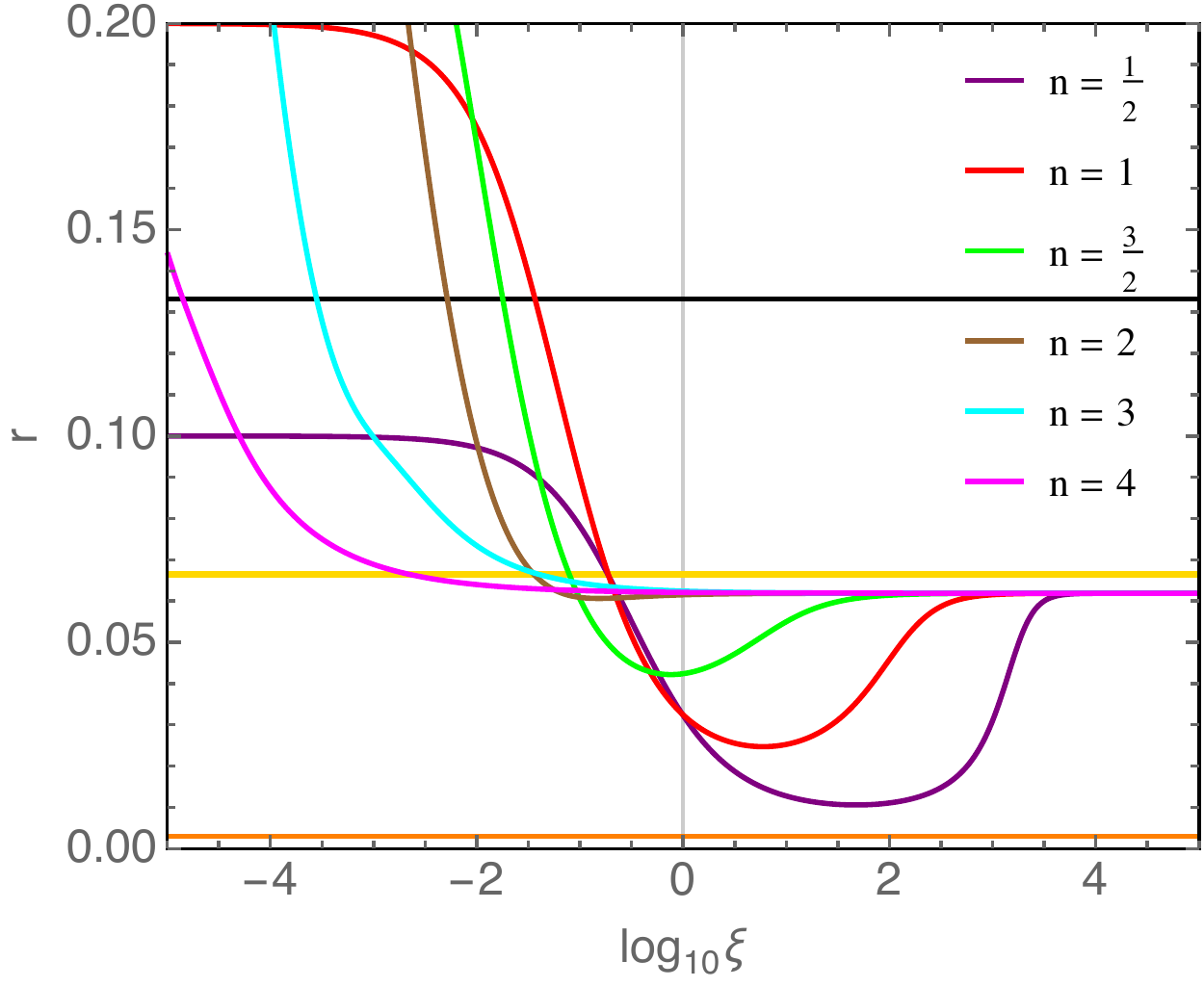}
\caption{Results of for the (\ref{eq:U2}) model with $f(\varphi) = 1 + \xi (\varphi/M_P)^{n}$. All solutions meet in the attractor point near to the result of linear inflation. {Around $\xi \sim 0.1$ at the right panel all of the curves almost coincide. This is purely accidental, since for the (\ref{eq:U2}) model the half of the trajectories reach the linear inflation attractor before $\xi \sim 0.1$.}
}
\label{fig:rvsnST}
\end{figure}

In this section we follow the effective field theory approach. Putting aside deep theoretical motivation, we  look for the class of models that would lead to linear inflation in a certain limit of the theory. From Eq. (\ref{eq:EF}), (\ref{eq:StrongCoupling}) one finds that in order to obtain $V \propto \phi$ in the strong coupling limit one requires
\begin{equation}
U = M^4 f^2 \, \log f \, , \label{eq:U1}
\end{equation}
{ where $M$ is a constant with dimension of a mass, obtained from the normalisation of primordial inhomogeneities \cite{Ade:2015lrj}
\begin{equation}
A_s = (2.14 \pm 0.05) \times 10^{-9}. \label{eq:As}
\end{equation}
}
In such a case the Einstein frame potential is proportional to $\log f$ and $f(\varphi)$ has to be always positive with a minimum at $f = 1$. The other option is to consider a Jordan frame potential, which asymptotically gives $V \propto \phi$ in the large coupling and $f \gg 1$ limits, for instance
\begin{equation}
U = M^4 (f-1)^2 \, \log f \, . \label{eq:U2}
\end{equation}
In the case of the model (\ref{eq:U2}), the Jordan frame theory has a potential similar to the Coleman-Weinberg model, which is proven to generate linear inflation in the large coupling limit \cite{Rinaldi:2015yoa,Kannike:2015kda}. The tensor-to-scalar ratio and scalar spectral index of models (\ref{eq:U1}) and (\ref{eq:U2}) for $N_\star = 60$ are plotted in Fig. \ref{fig:rvsnST2} and \ref{fig:rvsnST}. The straight continuous black/yellow/orange line represents quadratic/linear/Starobinsky inflation for number of e-folds between 50 and 60. We have assumed
\begin{equation}
f(\varphi) = 1 + \xi \left(\frac{\varphi}{M_P}\right)^{n} \, . \label{eq:non:min:f}
\end{equation}
where the inflaton, $\varphi$, is assumed to be a gauge singlet, therefore such choice does not break any symmetry of the SM or GR. For $2n$ odd values we assume that {due to some inner symmetry} the field value is constrained to positive values or equivalently $\varphi \to |\varphi|$. This ensures that the non-minimal coupling is always well defined and with a minimum at $\varphi =0$. The result characteristic for linear inflation is an attractor for all considered values of $n$.
Note that alike in the induced gravity scenario the value of $\xi$ for which the attractor solution is reached is inversely proportional to $n$.
{ Note that for $\xi \to 0$ the corresponding Einstein frame potential of eq. (\ref{eq:U1}) behaves like
\begin{equation}
 V \approx M^4 \xi \left(\frac{\varphi}{M_P}\right)^{n},
\end{equation}
reproducing linear inflation if $n=1$.
Moreover for $\xi \ll 1$ the attractor solution is also obtained in the $n = 2$ case. The same does not happen for the potential (\ref{eq:U2}) because for $\xi \to 0$ the corresponding Einstein frame potential of eq. (\ref{eq:U1}) behaves like $V \approx \varphi^{3n}$.}
In the model (\ref{eq:U2}) the attractor point (which is equivalent to the result of the $V \propto \phi (1-\exp(-\sqrt{2/3} \, \phi/M_P))^2$) EF potential) is slightly shifted from the result of linear inflation, due to the fact that the JF potential deviates from (\ref{eq:U1}). {Not surprisingly, the attractor is also $n$-independent, since in the strong coupling limit the EF potential does not depend on the form of $f(\varphi)$.}
\\*

This approach is similar to the induced gravity scenario presented in Ref. \cite{Kallosh:2013tua,Kallosh:2014laa,Giudice:2014toa,Kallosh:2014rha}. In the induced gravity inflationary scenario one wants to obtain a generalisation of Starobinsky inflation with a wide class of $f(\varphi)$ functions. To do so one assumes $U = M^4(f-1)^2$, which in the strong coupling limit gives the result of the Starobinsky inflation. Therefore our model can be considered as a modification of the induced gravity inflationary scenarios motivated by e.g. loop corrections to Jordan frame inflationary potential (see Section \ref{sec:QFT}).

\section{Particle Physics point of view} \label{sec:QFT}

\subsection{Explicit Planck Mass} \label{sec:QFT_MP}
{  From the quantum field theory point of view the tree-level inflaton potential $U$ in eq. (\ref{eq:JframeL}) cannot be an arbitrary function of $\varphi$ but it is \emph{usually} a polynomial expression. In analogy to what done in \cite{Kallosh:2013tua}, we focus on the case in which the scalar potential is just given by a single monomial term
\begin{equation}
U(\varphi) =  \lambda_{2n} M_*^{4-2n} \varphi^{2n}
 \, , \label{eq:UJ}
\end{equation}
where $2n$ is considered to be an integer value, $\lambda_{2n}$ is a dimensionless coupling and $M_*^{4-2n}$ is a dimensionful term needed to have the scalar potential with mass dimension equal to four. The constraint on the amplitude of scalar perturbations (\ref{eq:As})
allows us to fix only the product $\lambda_{2n} \, M_*^{4-2n}$. Therefore it is customary to take $M_*=M_P$ and then fix $\lambda_{2n}$ to an explicit numerical value \cite{Ade:2015lrj}.
 In case $n>2$, $M_*$ sets the cutoff energy scale of the theory. In this case, such kind of couplings can be generated by quantum gravity effects therefore it is even natural to take $M_*=M_P$. From now on this will be our choice. Moreover, $\lambda_{2n}$ is inevitably subject to quantum corrections because the inflaton must be coupled to some other particles in order to successfully reheat the universe \cite{Kannike:2015apa,
Kannike:2015kda,Marzola:2015xbh,Marzola:2016xgb} or by an underlying completion of the theory. The running of $\lambda_{2n}$ is described by its beta function
\begin{equation}
\beta (\mu) = \frac{d\lambda_{2n}}{d\log \mu} \, \label{eq:RGElambda}.
\end{equation}
where $\mu$ is the renormalization scale. Without knowing all the details of the inflaton interactions or of the theory completion, we can solve eq. (\ref{eq:RGElambda}) with a Taylor series
\begin{equation}
\lambda_{2n} (\varphi) = \lambda_{2n}(\mu_0) + \sum_{k=1}^\infty \frac{\beta_k}{k!} \log^k \left(\frac{\varphi}{\mu_0}\right) \label{eq:lambdafull}
 \, .
\end{equation}
where $\lambda_{2n}(\mu_0)$ is the value of $\lambda_{2n}$ at a scale $\mu_0$ and the $\beta_k$ parameters represent the $k$-th derivatives of the beta function computed at the scale $\mu_0$.
Assuming that we obtain a good approximation of the expansion in (\ref{eq:lambdafull}) by keeping only the first order correction, we can reparametrize the running $\lambda_{2n} (\varphi)$ as
\begin{equation}
\lambda_{2n} (\varphi) \simeq \beta_1  \log\left(\frac{\varphi}{\varphi_0}\right) \label{eq:lambda}
 \, .
\end{equation}
where $\varphi_0= \mu_0 \exp\left( -\frac{\lambda_{2n}(\mu_0) }{ \beta_1} \right)$.
}
Therefore we can write the 1-loop inflaton potential as\footnote{{ It has been shown that the cosmological perturbations are invariant under
a change of frames (see for instance \cite{Prokopec:2013zya,Jarv:2016sow} and references therein), however the quantum level equivalence of the Jordan and the Einstein
frames is still an open issue. For example in \cite{Kamenshchik:2014waa} it has been shown that the one-loop divergences induce an off-shell frame dependence. In the present article our computation strategy is the following: we consider the leading order loop corrections to the Jordan frame scalar potential to be eq. (\ref{eq:UeffJ}). Once we have reached the full expression of the 1-loop Jordan frame scalar potential, we move to the Einstein frame in order to have a more straightforward computation of the slow-roll parameters. Given a scalar potential expression in the Jordan frame, the cosmological perturbations are then independent, in the slow-roll approximation, from the choice of the frame in which we perform the computation of the inflationay parameters \cite{Prokopec:2013zya,Jarv:2016sow}.} For other ways to include a loop correction in a scalar-tensor theory see Refs. \cite{Bezrukov:2010jz,George:2013iia,George:2015nza,Miao:2015oba,Barvinsky:2008ia,Bezrukov:2008ej,DeSimone:2008ei,Barvinsky:2009fy,Steinwachs:2011zs}.}
\begin{equation}
U_\text{eff} \simeq \beta_1 \, M_P^{4-2n} \, \log\left(\frac{\varphi}{\varphi_0}\right) \varphi^{2n}
 \, .
\label{eq:UeffJ}
\end{equation}
and we assume that any other loop contribution is subdominant. We want to stress that such a form of the Jordan frame potential is valid only in the high energy regime\footnote{This can be easily achieved if the quantum corrections are generated by a massive particle: the loop corrections are active only for scales higher than such particle mass.}, i.e. around the vacuum of the Jordan frame potential one restores the tree-level potential in Eq. (\ref{eq:UJ}).

After having defined the inflaton potential, let us discuss its coupling with gravity. We suppose the same non-minimal coupling function as in eq. (\ref{eq:non:min:f}). In analogy with what done in \cite{Kannike:2015apa,Kannike:2015kda,Marzola:2015xbh,Marzola:2016xgb}, we assume that $\xi$ does not receive relevant quantum corrections.
{ Generally the non-minimal coupling is subject to quantum corrections parametrized by a beta-function of the following type
\begin{equation}
  16 \pi^2 \beta_{\xi} \approx \xi \sum_k \lambda_k  .
  \label{eq:betaxi}
\end{equation}
where $\sum_k \lambda_k$ represents some other couplings from the scalar potential (for example the ones that are also generating the running of $\lambda_{2n}$) that right now we are not going to specify. In order to ignore the quantum corrections to $\xi$, the condition $\beta_{\xi} \ll \xi $ must be satisfied. This has been explicitly realized for $n=2$ in \cite{Kannike:2015apa,Kannike:2015kda,Marzola:2015xbh,Marzola:2016xgb}.
However, because of the constraint on the amplitude of scalar perturbations (\ref{eq:As}),
perturbativity of the theory and the $16 \pi^2$ suppression factor, we assume that such condition may be realized also for the other values of $n$.
}Taking now the strong coupling limit, the non-minimal coupling function and the Einstein field value are respectively approximated as:
\begin{equation}
f(\varphi) \simeq \xi \left(\frac{\varphi}{M_P}\right)^{n}  \, , \qquad
\phi \approx M_P \log\left(\frac{\varphi}{M_P}\right) \, .
\end{equation}
Therefore the Einstein frame potential can be approximated as
\begin{equation}
V = \frac{U_\text{eff}}{f^2} \simeq \beta_1 \, M_P^{4-2n} \, \log\left(\frac{\varphi}{\varphi_0}\right) \frac{\varphi^{2n}}{\xi^2 \varphi^{2n}}  \approx \phi  \, .
\end{equation}

\begin{figure}[t]
\centering
\includegraphics[height=5cm,bb=0 0 360 293]{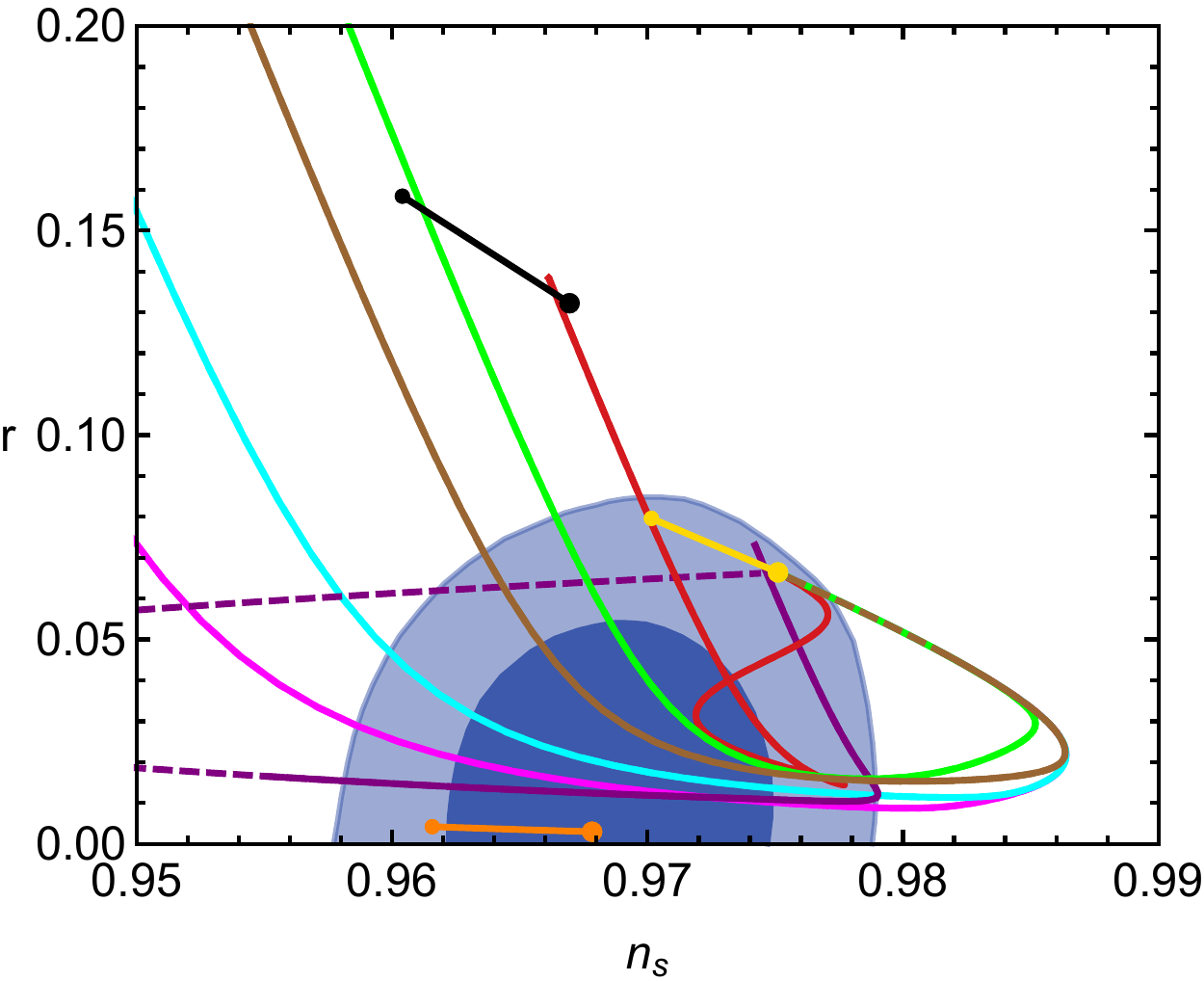}
\hspace{0.5cm}
\includegraphics[height=5cm,bb=0 0 360 303]{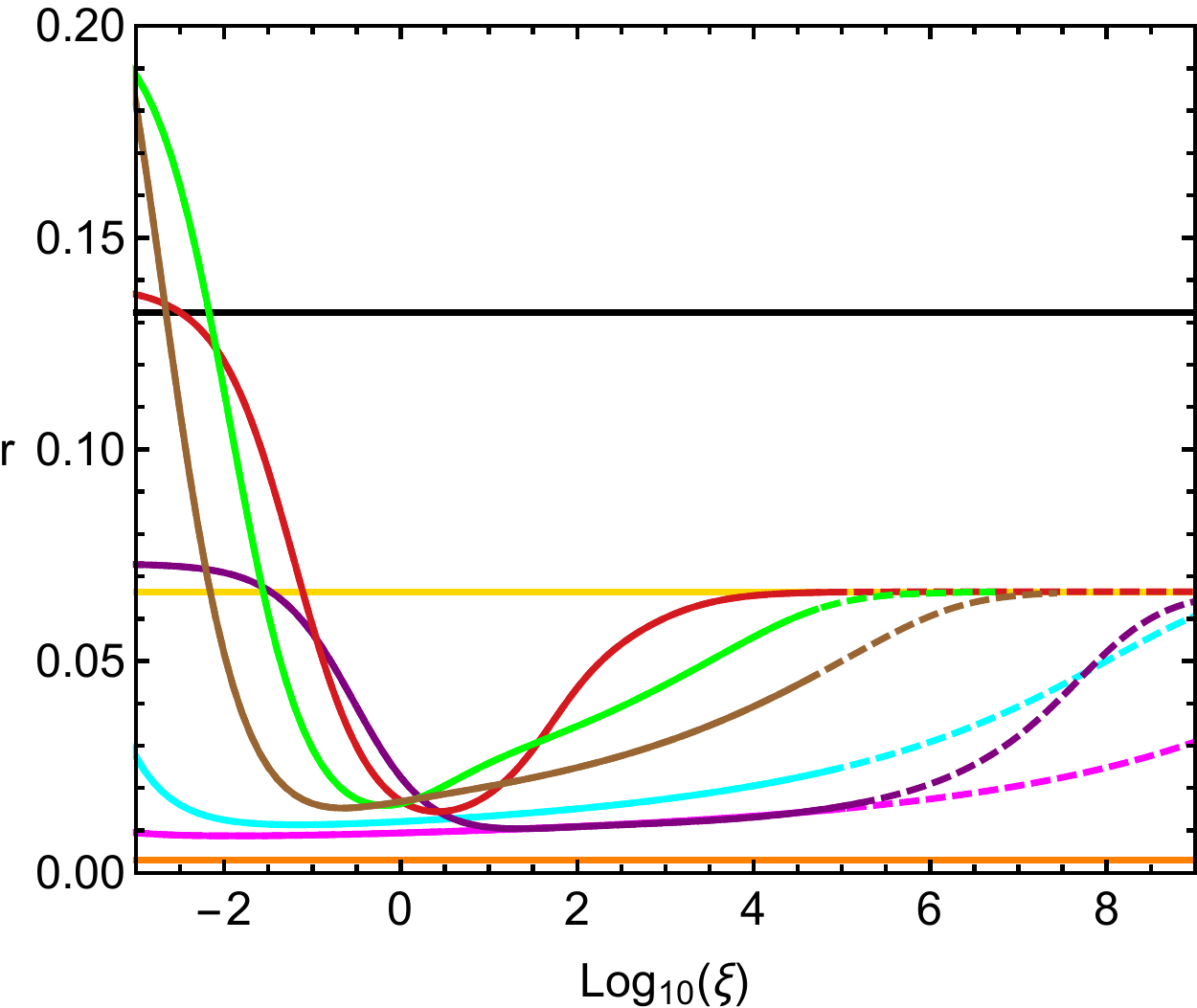}
\caption{Results of the particle physics motivated scenarios with $f(\varphi) = 1 + \xi \varphi^{n}$. {Note that unlike in the scalar-tensor scenario the smallest $\xi$ needed in order to reach the attractor is in the $n=1$ scenario.}}
\label{fig:rvsn}
\end{figure}

As we can see, the attractor solution is independent on the value of $n$. In the following { we consider the cases  $n=1/2,1,3/2,2,3,4$ and compare them with the results of \cite{Kallosh:2013tua}. The results are given in Fig. \ref{fig:rvsn} for the reference values of $60$ for the number of e-folds and $\varphi_0=10^{-3} M_P$.} The purple/red/green/brown/cyan/magenta line represents { $n=1/2,1,3/2,2,3,4$}, where continuous line represents $\beta_1 <1$ (the nominal perturbativity bound that allows to look for a completion of the theory in a perturbative way at scales higher than the inflation one) while the dashed line stands for $\beta_1 > 1$.
{ We can see in all cases that first, for smaller $\xi$, the predictions are aligned with the strong-coupling limit of the standard (without loop corrections) non-minimal inflation \cite{Kallosh:2013tua}, then, for higher $\xi$ values, the loop correction becomes relevant and the results are departing from the Starobinsky attractor and approaching the linear limit. As a consequence of this behaviour all the cases have a minimal value for the tensor-to-scalar ratio $r$. Such minimum depends on the $n$ value and on the exact value of $\varphi_0$. For given $n$, changing $\varphi_0$ will only affect the exact position of the minimum but the general behaviour and shape of the $(r,\ n_s)$ plot remains unchanged. Therefore we plot only the $\varphi_0=10^{-3} M_P$ case as an illustrative one.}
We also notice that the quantum corrected potentials are reaching the linear attractor for bigger $\xi$ values than the ones needed by the corresponding tree-level potentials to reach the Starobinsky attractor  \cite{Kallosh:2013tua}. However, only the case $n=1$  reaches the linear attractor solution inside the perturbative limit, making this particular case worthy of further investigation, which is postponed to a forthcoming work. { Moreover we stress that our results differ from Higgs-inflation with loop corrections studied in \cite{Hamada:2014iga,Bezrukov:2014bra}, because in those scenarios, thanks to the tuning of the top mass, inflation is realized at the inflection point of the potential and eq. (\ref{eq:lambda}) is not anymore valid but higher order terms are needed in the $\lambda_{2n}$ expansion (see for instance eq. (4) of \cite{Hamada:2014iga}).}

Let us note that the case of $f = 1 + \xi (\varphi/M_P)^{n}$ may be fully consistent with the scalar-tensor approach. In such a case the JF potential (\ref{eq:U2}) has a stable GR minimum, which also provides the graceful exit. The main difference appears within the $\log f$ term, which also can be expressed in the form similar to $\log(\varphi/\varphi_0)$. For $f = 1 + \xi (\varphi/M_P)^{n} \gg 1$ one finds
\begin{equation}
\log f \simeq \log\left(\xi (\varphi/M_P)^{n}\right) = n\log\left(\frac{\varphi}{\varphi_0}\right) \, ,
\end{equation}
where $\varphi_0 = M_P \,  \xi^{-n}$. In such a case the effective $\varphi_0$ is directly related to the strength of the non-minimal coupling to gravity. Therefore assuming $\varphi_0 =  10^{-3}$ $M_P$ one finds $\xi \simeq 10^{6/n}$. The relation between $\xi$ and $\varphi_0$ can be changed for different forms of the JF potential. For instance for $U = M^4 (f-1)^2\log(f/f_0)$ (where $f_0 \propto \xi$) one can keep $\varphi_0$ as a $\xi$-independent parameter, just like in the particle physics approach. Note that in both, ST and QFT approaches, our model is simply the induced gravity inflation with additional log term, which may be originated from loop corrections. Loop corrections appear naturally in all inflationary models, affecting the flatness of the potential \cite{Kannike:2014mia,Marzola:2015xbh,Marzola:2016xgb,Kannike:2015apa,
Kannike:2015kda,Kannike:2016jfs,Rinaldi:2015yoa,
Elizalde:2015nya,Inagaki:2015fva,Ballesteros:2015noa,Grozdanov:2015zna}. Therefore one of the main conclusion is that as long as there are no symmetries\footnote{The inflaton could be for instance a Goldstone boson or be protected by e.g. shift symmetry or supersymmetry in order to keep the loop corrections small. {It has has been pointed out in \cite{Bezrukov:2010jz,Barvinsky:2009ii} that the ``duality'', between an approximate scale invariance in the Jordan frame and a shift symmetry in the Einstein frame, may protect the
effective potential from dangerous loop corrections. However in \cite{Kannike:2015apa,Kannike:2015kda,Marzola:2015xbh,Marzola:2016xgb} it has been explicitly shown that even scale invariance might not protect from such loop corrections.}}, which would protect the potential, loop corrections should move the attractor point of induced gravity from Starobinsky model to linear inflation.

\subsection{Dynamically Induced Planck Mass}

\begin{figure}[t]
\centering
\includegraphics[height=5cm,bb=0 0 360 293]{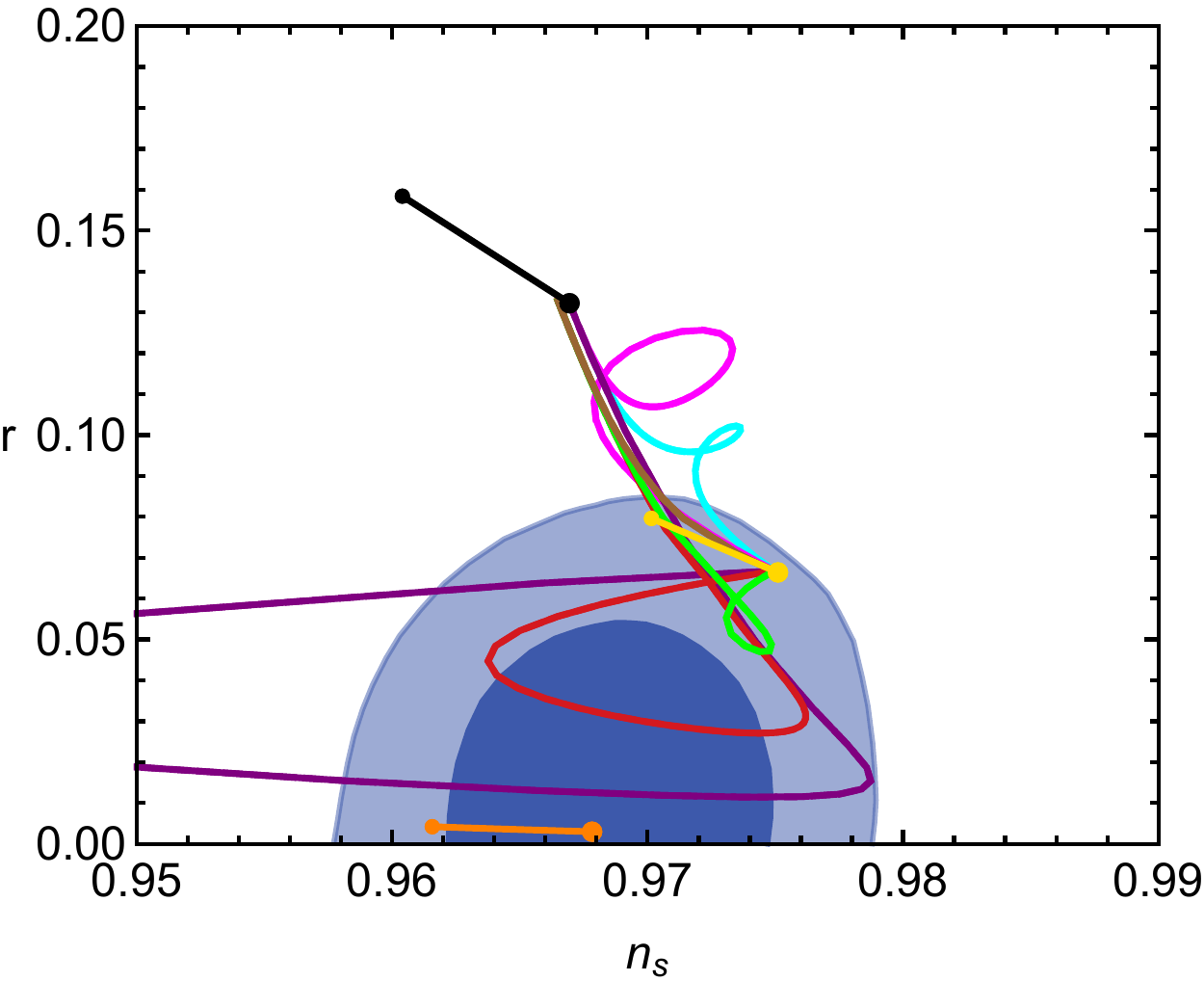}
\hspace{0.5cm}
\includegraphics[height=5cm,bb=0 0 360 303]{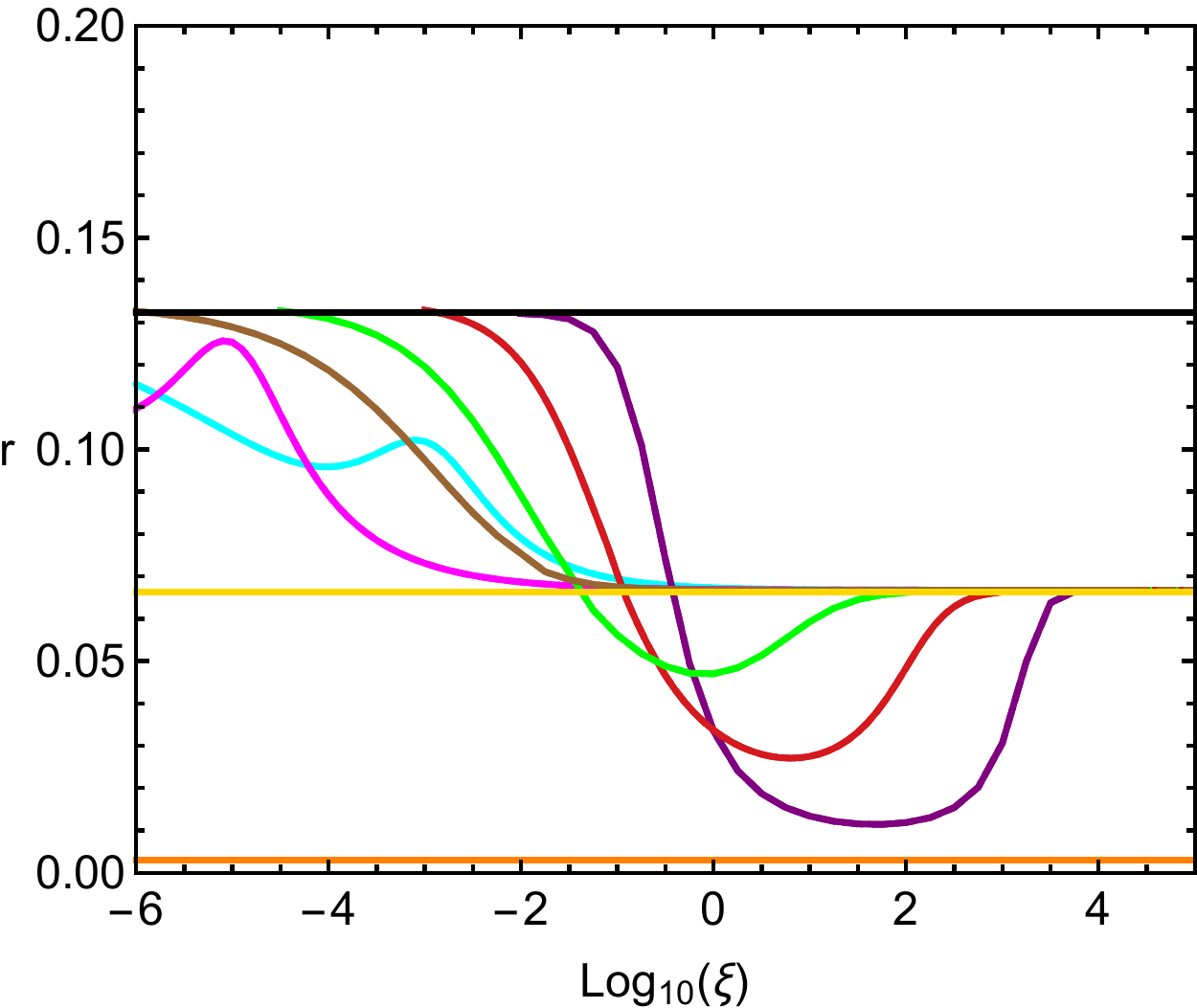}
\caption{Results of the particle physics motivated scenarios with $f(\varphi) = \xi \varphi^{n/2}$. {Note that now the attractor solutions are reached always in the perturbative region.}}
\label{fig:rvsn_ind}
\end{figure}

As another possibility we consider a scenario in which the approximated potential in Eq. (\ref{eq:UeffJ}) is valid also at low energies\footnote{This can be easily achieved if the quantum corrections are generated by a massless particle, for instance a dark photon.} i.e. the tree-level potential in Eq. (\ref{eq:UJ}) is not restored. In this case then, $\varphi$ acquires a VEV different from zero, $v_\varphi = e^{-\frac{1}{2n}} \varphi _0$. In such a case, in order to get a proper minimum of the potential, $U_\text{eff}(v_\varphi)=0$, we must add a cosmological constant\footnote{In the scalar-tensor approach such a situation is avoided because the argument of the $\log$ is not simply the field value $\varphi$ but $f(\varphi)=1+\xi (\varphi/M_P)^n$, so that in the minimum $\log(f(0))=0$.} to the potential so that
\begin{equation}
U_\text{eff} \simeq \beta_1 \, M_P^{4-2n} \, \log\left(\frac{\varphi}{\varphi_0}\right) \varphi^{2n} + \Lambda^4
 \, ,
\label{eq:UeffJ_2}
\end{equation}
where
\begin{equation}
\Lambda = \sqrt[4]{\frac{\beta_1}{2 \, e \, n}} M_P^{1-n/2} \, \varphi_0^{n/2}
 \, .
\label{eq:Lambda}
\end{equation}
After having defined the inflaton potential, let us discuss its coupling with gravity. In analogy with \cite{Kannike:2015kda,Kannike:2015apa}, we assume that the Planck mass is dynamically generated in the Jordan frame by the inflaton VEV  $v_\varphi$, therefore
\begin{equation}
f(\varphi) = \xi \left(\frac{\varphi}{M_P}\right)^{n} \quad \text{ and } \quad \xi \left(\frac{v_\varphi}{M_P}\right)^{n} = 1
 \, .
\end{equation}
Then we can repeat the same procedure as illustrated in section \ref{sec:QFT_MP}. The results are given in Fig. \ref{fig:rvsn_ind}. The colour code is the same as in Fig. \ref{fig:rvsn}. We can see that now the attractor solutions are reached faster for higher $n$ values and always in the perturbative region. Moreover we notice that for small $\xi$ we always get quadratic inflation results. This can easily explained  taking the limit of the Einstein frame potential for $\xi$ going to zero:
\begin{equation}
V(\phi)_{\xi \to 0} \approx \beta_1 \, M_P^2 \, n  \, \xi ^{\frac{2}{n}-2} \, \phi ^2
 \, ,
\end{equation}
in agreement with the results in Fig. \ref{fig:rvsn_ind}. { Finally we notice that, in correspondence with the curl in the $r$ vs. $n_s$ plot, for $n< (>)2$ the $r$ vs. $\xi$ plot exhibits a local minimum (maximum). The $n=2$ case present no curl in the $r$ vs. $n_s$ plot (and therefore no corresponding stationary point in the $r$ vs. $\xi$ plot) due to the approximate classical scale symmetry of the scalar potential broken by only $\Lambda$ and the running of $\lambda_{2n}$.}


\section{Summary} \label{sec:Summary}

In this work we presented two ways of obtaining a linear inflation attractor in a scalar-tensor theory. In Sec. \ref{sec:general} we investigate a theory with a non-minimal coupling to gravity of the form $M_P^2 f(\varphi)R$. We note that for the JF scalar potential $U = M^4 f^2 \log f$ or $U = M^4 (f-1)^2 \log f$, one obtains in the strong coupling limit the EF potential proportional to the EF field. We investigate the $f = 1 + \xi (\varphi/M_P)^n$ scenario and we prove, that for sufficiently big $\xi$ one always restores the results of linear inflation. This attractor solution may be even reached for $\xi \ll 1$ for $n\geq 2$. {For specific forms of $f(\varphi)$ the attractor may be far from the linear inflation result. For instance, if $f$ contains a stationary point, inflation happens in a weak coupling regime. In the case of infinite order stationary point the linear inflation attractor cannot be obtained for any values of $\xi$ (see appendix \ref{sec:App} for details).}
\\*

In Sec. \ref{sec:QFT} we show that the $\log$ term may naturally be motivated by the loop corrections in quantum field theory. In this approach the $\log$ term does not contain the $f(\varphi)$ function. For both, explicit and induced Planck mass we have obtained an attractor on the $(n_s,r)$ plane, which corresponds to the result of linear inflation. We have also discussed under which circumstances one obtains the equivalence between ST and QFT approaches. Our model is simply an induced gravity inflationary scenario with a loop correction. Therefore (assuming the loop correction cannot be neglected) linear inflation is a new attractor solution for induced gravity inflation.

\acknowledgments

MA would like to thank Fedor Bezrukov and Apostolos Pilaftsis for useful discussion. MA was supported by the Iuventus Plus grant No. 0290/IP3/2016/74 from the Polish Ministry of Science and Higher Education. AR was supported by the Estonian Research Council grants IUT23-6, PUT1026 and by the ERDF CoE grant TK133.

\begin{appendix}

\section*{Appendix}

\section{Linear inflation and local flatness} \label{sec:App}

To avoid the linear inflation attractor one needs to obtain a flat region of the EF potential beyond the strong coupling limit. As shown in \cite{Artymowski:2016ikw} such a flat region may appear in the induced gravity inflation as long as $f(\varphi)$ has a stationary point of order $m-1$ at some $\varphi_s$. Then $\varphi_s$ becomes a stationary point of the EF potential, which obtains another flat region in vicinity of $\varphi_s$. Assuming
\begin{equation}
f(\varphi) = \xi \sum_{k=0}^m \, \lambda_k \, \left(\frac{\varphi}{M_P}\right)^k \, ,\label{eq:lambdak}
\end{equation}
the requirement of the existence of the stationary point indicates that
\begin{equation}
f(\varphi) = \frac{\xi}{m}\left(m \, \lambda\right)^{\frac{-1}{m-1}}\left(1 + \left(\left(m \, \lambda\right)^{\frac{1}{m-1}}\frac{\varphi}{M_P} - 1 \right)^m\right) \, , \quad \varphi_s = M_P \left(m \, \lambda\right)^{\frac{-1}{m-1}}\, ,\label{eq:fgeneral}
\end{equation}
where $\lambda:= \lambda_m$ is a free parameter of the theory. The crucial point is that $\varphi_s$ is always in the weak coupling regime, where the $\log$ term does not generate the linear inflation attractor. For finite $m$ one obtains two phases of inflation, namely linear inflation for $\varphi/\varphi_s \gg 1$ and a stationary-point inflation for $\varphi \sim \varphi_s$. In such a case linear inflation is a form of a proto-inflation, which homogenise the Universe before the inflation around the stationary point, solving the problem of initial conditions for inflation described in the Ref. \cite{Ijjas:2013vea}. Note that in the $m\to \infty$ limit (for which $f = 2 - e^{-\xi \, \varphi/M_P}$) inflation can only happen in the weak coupling regime, i.e. the $M_P^2 \, f_\varphi^2 \gg f$ condition is not satisfied for any $\xi$. Therefore the $\log$ term is negligible for all values of $\xi$. This idea will be further explored in a forthcoming work.

\end{appendix}

\bibliographystyle{JHEP}
\bibliography{citations}

\end{document}